\documentstyle[12pt,prl,aps]{revtex} 
\large
\def\addcontentsline#1#2#3{\relax}

\long\outer\def\demo#1. #2\par{\medbreak\noindent {\bf#1.\enspace}
        {\rm#2}\par\ifdim\lastskip<\medskipamount\removelastskip
        \penalty55\medskip\fi}
\newcommand{\ben}{\begin{enumerate}}
\newcommand{\een}{\end{enumerate}}
\def\bdemo#1. #2\par{\medbreak\noindent {\bf#1.\enspace}{\rm#2}\par}
\def\edemo{\ifdim\lastskip<\medskipamount\removelastskip\penalty55\medskip\fi}

\def\0{{\bf 0}}
\def\1{{\bf 1}}

\def\c{{\bf c}}
\def\n{{\bf n}}
\def\x{{\bf x}}
\def\N{{\bf N}}
\def\Q{{\bf Q}}
\def\R{{\bf R}}
\def\D{{\rm D}}
\def\Exp{{\rm Exp}}
\def\diag{{\rm diag}}
\def\BD{\mathop{\rm BD}\nolimits}
\def\mod{\mathop{\rm mod}\nolimits}

\def\longto{\mathop{\longrightarrow}\limits}
\def\bfa{\mathop{a\kern-.5em{a}\kern-.5em{a}}\nolimits}
\def\bfb{\mathop{b\kern-.47em{b}\kern-.47em{b}}\nolimits}
\def\bfx{\mathop{x\kern-.6em{x}\kern-.6em{x}}\nolimits}
\begin{document}
\baselineskip 18pt
\author{Yshai Avishai$^{\,1,4}$, Dani Berend$^{\,2}$, 
and Richard Berkovits$^{\,3}$}
\address{${}^{1}$Department of Physics, Ben Gurion University 
of the Negev,Beer-Sheva 84 105, Israel \\
${}^{2}$Department of Mathematics and Computer Sciences,
 Ben Gurion University of the Negev,
Beer-Sheva 84 105, Israel \\
${}^{3}$Minerva Center, Department of Physics, Bar Ilan University,
Ramat Gan, Israel \\ 
${}^{4}$NTT Basic Research Laboratories, 3-1 Morinosato, Wakamiya,
Atsugi-shi Kanagawa-ken, Japan \\ }

\title{Fluctuation of inverse compressibility for electronic 
systems with random capacitive matrices}

\maketitle

\begin{abstract}
This article is concerned with statistics of addition spectra
of certain many body systems of identical particles. 
In the first part, the pertinent system consists 
of $N$ identical particles distributed among $K<N$
independent sub-systems, such that the energy of each
sub-system is a quadratic function of the number of
particles residing on it with 
random coefficients. 
On a large scale, the ground state energy 
$E(N)$ of the whole system grows
quadratically with $N$, but in general there is no simple
relation such as $E_{N}=a N + b N^{2}$. 
The deviation of $E(N)$ from exact quadratic 
behavior implies that its second 
difference (the inverse compressibility) 
$\chi_{N} \equiv E(N+1)-2 E(N) 
+E(N-1)$ is a fluctuating quantity.
Regarding the numbers $\chi_{N}$ as values assumed by a certain random
variable $\chi$, we obtain a closed-form expression for 
its distribution $F(\chi)$. Its main feature is 
that the corresponding density $P(\chi)=\frac{dF(\chi)} {d\chi}$ 
has a maximum at the point $\chi=0$. As $K \rightarrow \infty$ 
the density is Poissonian, namely, $P(\chi) \rightarrow e^{-\chi}$.\\

This result serves as a starting point for the second part 
in which coupling between subsystems is included. 
More generally, a classical model is suggested
in order to study
fluctuations of Coulomb blockade peak spacings 
in large two-dimensional semiconductor quantum dots.
It is based on the electrostatics 
of several electron islands among which there are
random inductive and capacitive couplings.
Each island can accommodate 
electrons on quantum orbitals whose 
energy depend also on an external magnetic field. 
In contrast with a single island 
quantum dot where the spacing distribution
between conductance peaks is close to Gaussian,
here the distribution has a peak at small 
spacing value. The fluctuations are mainly due 
 to charging effects. 
 The model 
can explain the occasional occurrence of 
couples or even triples of closely spaced Coulomb blockade peaks, 
as well as the qualitative behavior of peak positions 
with the applied magnetic field. 
 \\

\end{abstract}


\newpage
\section {Statistics of addition spectra of independent quantum systems}
\subsection {Motivation}
Statistics of spectra is an efficient tool for elucidating 
properties  of various physical systems. So far, most of 
the effort is focused on the study of energy levels of a system
with a fixed number of particles. In this 
context, one of the central earlier results 
is that the spectral statistics of many-body 
systems such as complex nuclei agree with the predictions of 
random matrix theory \cite{Wigner,Dyson}. On the 
other extreme, it was found that level statistics of 
a single particle in a chaotic or disordered system also 
obeys a Wigner-Dyson statistics \cite{Bohigas,Altshuler}.

Recently, experiments are designed to get information on 
the statistics of the {\em addition spectra} of electrons 
in quantum dots \cite{Sivan}. The pertinent energy levels
$E(N)$ are the ground state energies of a system consisting of $N$ electrons 
residing on a quantum dot, which is coupled capacitively to its environment.

Let us single out two properties of the addition 
spectra of quantum dots. The first one is that, on a large scale, 
the energy $E(N)$ grows quadratically with $N$, while the second one is 
a consequence of charge quantization, namely, there is, in general, no simple 
relation such as $E(N)=a N +b N^{2}$. In this context, an appropriate quantity 
whose statistics is of interest is then the inverse compressibility,
\begin{eqnarray}
\chi_{N} \equiv E(N+1)-2 E(N)+E(N-1).
\label{eq_compress}
\end{eqnarray}
It is the deviation of $E(N)$ from exact quadratic behavior which makes its
second difference $\chi_{N}$ non-constant. Indeed,
in a recent experiment on large quantum dot \cite{ZAPW} it was found that the
inverse compressibility vanishes for numerous values of electron number~$N$.

In the present section we study the statistics of the addition spectrum of a
simple physical system with the two basic properties mentioned above. One 
example of such a system is motivated by considering the electrostatic energy
of large quantum dots (although it should be mentioned that the model is too
simple to describe the actual physics, see next section).
To be specific, we have in mind a system of $K$ metallic 
grains such that the number of electrons on the $i^{th}$
grain is $n_{i}$ ($i=0,1,2,\ldots,K-1$) and their sum equals $N$. The
electrostatic energy of the pertinent system is a bilinear form in the 
numbers $n_{i}$ with a $K \times K$ matrix $w \equiv \frac {1} {2} C^{-1}$.
Here $C$ is a positive-definite symmetric matrix of 
capacitance and inductance coefficients.
If the metallic grains are very far apart, the matrix $C$ is nearly diagonal.
Thus, we concentrate on the special case $C=\diag[C_{i}]$, for which the energy
of the system is given by 
\begin{eqnarray}
E(N)=\min \sum_{i=0}^{K-1} \frac {1} {2 C_{i}}
n_{i}^{2}, \mbox {(subject to $\sum_{i=0}^{K-1} n_{i}=N).$}
\label{eq_EN}
\end{eqnarray} 
The minimum in (\ref{eq_EN}) is taken 
over all possible partitions $n_i$ of $N$.

Another example is the energy of a system composed of $K$ different harmonic
oscillators, among which one distributes $N$ spinless fermions.
If there are $n_{i}$ fermions on 
oscillator $i$ (whose frequency is $\omega_{i}$),
then the energy of this oscillator (up to a constant) is
$E_{i}=\hbar \omega_{i} n_{i} (n_{i}+1)$,
and hence the ground state energy of the system is
\begin{eqnarray}
E(N)=\min \sum_{i=0}^{K-1} E_{i},
\mbox {(subject to $\sum_{i=0}^{K-1} n_{i}=N).$}
\label{eq_ENOS}
\end{eqnarray}
Discussion is concentrated 
on the first example, which is borrowed from
the electrostatics of quantum dots (\ref{eq_EN}), and 
refer to the constants $C_{i}$ as capacitors. A
remark pertaining to the
second example (the system of oscillators 
(\ref{eq_ENOS})) are also presented.

Regarding the numbers $\chi_{N}$ of (\ref{eq_compress}) as values assumed by a 
certain random variable, the distribution of this random variable is the main
focus of the present work, which culminates in Theorem~1,
where we find a closed-form expression for the distribution.

The problem of elucidating the (addition) spectral statistics of a
{\em a many-body} system, consisting of several independent sub-systems 
(whose dependence of $E$ on $n_{i}$ is known), looks deceptively simple.
As will be evident shortly, this is not the case, and finding 
the distribution in question is quite a non-trivial task. Note that,
even for a {\em single particle system} composed of several 
independent sub-systems ({\em e.g.,} a system of a particle in several boxes),
the derivation of level statistics requires a large degree 
of mathematical effort \cite{Berry}. The rest of this section 
is therefore devoted
to a rigorous derivation of our main results. 
We offer our apology to the causal 
reader who might find this part rather mathematically oriented. 
He or she is adviced to start looking at {\bf problem 1} 
below and leave this section after 
reading the statement of {\bf Theorem 1}.
\subsection{Formalism}
\demo Definition 1. Let $(\theta_n)_{n=1}^\infty$ be a sequence of real numbers
and $F$ a distribution function. The sequence $(\theta_n)$ is {\it asymptotically
$F$-distributed} if
$$\frac{\left|\{1\le n\le M:\theta_n\le x\}\right|} {M}\longto_{M\to\infty} F(x)$$
for every continuity point $x$ of $F$ (where $|S|$ denotes the cardinality of
a finite set $S$).

An equivalent condition is the following. Denote by $\delta_t$ the point mass
at $t$, and let $\mu$ be the probability measure corresponding to the distribution
$F$ (namely, $\mu(A)=\int 1_A dF(x)$ for any Borel set $A$). Then $(\theta_n)$ is
asymptotically $F$-distributed if
$$\frac{1}{M}\left(\delta_{\theta_1}+\delta_{\theta_2}+\ldots+\delta_{\theta_M}\right)
\longto_{M\to\infty} \mu$$
(the convergence being in the weak*-topology).

The notion of asymptotic distribution has a stronger version whereby, instead of
requiring only that initial
pieces of the sequence behave in a certain way, we require this to
happen for any large finite portion of the sequence. This leads to

\demo Definition 2. In the setup of Definition 1, $(\theta_n)$ is
{\it asymptotically well $F$-distributed} if
$$\frac{\left|\{L< n\le M:\theta_n\le x\}\right|} {M-L}\longto_{M-L\to\infty} F(x)$$
for every continuity point $x$ of $F$.

Recall that the {\it density} of a set $A\subseteq\N$ is given by
$$\D(A)=\mathop{\lim}_{M\to\infty} {\left|A\cap [1,M]\right|\over M}$$
if the limits exists. If, moreover, the limit
$$\BD(A)=\mathop{\lim}_{M-L\to\infty} {\left|A\cap (L,M]\right|\over M-L}$$
exists (in which case it is certainly the same as $\D(A)$), then it is called
the {\it Banach density} of $A$ (cf. \cite[p.72]{Fu}).

The following lemma is routine.

\proclaim Lemma 1. Let $(\theta_n)_{n=1}^\infty$ be a sequence of real numbers.
Suppose $\N=\bigcup_{j=1}^r A_j$, where the union is disjoint. Let
$(\theta^{(j)}_n)_{n=1}^\infty$ be the subsequence of $(\theta_n)$, consisting of
those elements $\theta_n$ with $n\in A_j,\ 1\le j\le r$.
\ben
\item
If each $(\theta^{(j)}_n)$ is asymptotically $F_j$-distributed for some
distribution functions $F_j,\ 1\le j\le r$, and $\D(A_j)=d_j,\ 1\le j\le r$, then
$(\theta_n)$ is asymptotically $F$-distributed, where $F=\sum_{j=1}^r d_j F_j$.
\item
If each $(\theta^{(j)}_n)$ is asymptotically well $F_j$-distributed and
$\BD(A_j)=d_j$, then $(\theta_n)$ is asymptotically well $F$-distributed.
\een

Obviously, a general sequence on the line does not have to be asymptotically
distributed according to some distribution function, but one would expect it of
sufficiently ``regular" bounded sequences. In our case, one might expect $\chi_{N}$
to be distributed according to some distribution function corresponding to a measure
centered at about $1/C$. However, this is not the case. In fact, the measure in
question is supported on a finite interval, and is a convex combination of an
absolutely continuous measure with decreasing density function on some interval
$[0,a]$ and the point mass $\delta_a$ at the right end $a$ of that interval.

We have defined $E(N)$ indirectly by means of the following

{\bf Problem 1}. For each non-negative integer $N$, find non-negative integers
$n_0,n_1,\ldots,n_{K-1}$, satisfying $n_0+n_1+\ldots+n_{K-1}=N$, for which
$\sum_{i=0}^{K-1} \frac{1}{2C_i}\cdot n_i^2$ is minimal.

It turns out that this problem is intimately related to a second optimization problem.
Put $w_i=\frac{1}{2C_i},\ 0\le i\le K-1$, and let $\Delta$ denote the set of all
positive odd multiples of the numbers $\frac{1}{2C_i}$:
$$\Delta=\{w_0,3w_0,5w_0,\ldots,w_1,3w_1,5w_1,\ldots,
           w_{K-1},3w_{K-1},5w_{K-1},\ldots\}\;.$$
Here we treat $\Delta$ as a multi-set, or a sequence, in the sense that if some elements
appear in this representation of $\Delta$ more than once (which occurs iff some ratio
$w_i/w_j$ is a rational number with odd numerator and denominator), then we
consider $\Delta$ as having several copies of these numbers.

{\bf Problem 2}. For each non-negative integer $N$, minimize
$\sum_{m=1}^N \delta_m$, where $\delta_1,\delta_2,\ldots,\delta_N$ range over all distinct
$N$-tuples in $\Delta$.

Note that, if an element appears several times in $\Delta$, it is allowed to appear the
same number of times in the sum as well.

Let us demonstrate the equivalence of the two problems. Given the sum
$\sum_{i=0}^{K-1} w_i\cdot n_i^2$, we may use the equality
$w_i\cdot n_i^2=w_i+3w_i+5w_i+\ldots+(2n_i-1)w_i$
to see that any feasible value for the objective function of the first problem is a
feasible value for the objective function of the second problem as well. On the other
hand, solving Problem 2 is trivial. Namely, one minimizes the sum there simply by
taking the $N$ least elements of the set $\Delta$. In particular, for each $i$, the
multiples of $w_i$ present in the optimal solution will be all odd multiples
$w_i,3w_i,5w_i,\ldots$ up to some $(2n_i-1)w_i$. Thus, the optimal solution of
Problem 2 yields the optimal solution of Problem 1 also. We note in passing that
this discussion shows also that the minimum (for each of the problems) is obtained
at a unique point unless $\Delta$ contains multiple elements. (However, we shall
always refer to {\bf the} optimal solution, even when there may be several.)

A simple consequence of the above is

\proclaim Proposition 1. Let $\n=(n_i)_{i=0}^{K-1}$ be the optimal solution
of Problem 1 for some value of $N$. Then the optimal solution of Problem 1,
with $N+1$ instead of $N$, is $\n'=(n'_i)_{i=0}^{K-1}$, where $n'_j=n_j+1$ for
some $0\le j\le K-1$ and $n'_i=n_i$ for $i\ne j$.

\demo Remark. It is convenient to comment here on the effect a certain change
in the original problem would make. One may consider the energies $E_i$ to
be $w_i n_i(n_i+1)$ instead of $w_i n_i^2$. This would change $\Delta$ to be
the set of all even multiples of the $w_i$'s. Obviously, this would leave intact the
equivalence of Problems~1 and 2. One can check that this would have also no effect
on Theorems~1 and~2 {\it infra}.

To formulate our main result we need a few definitions and notations. Real numbers
$\theta_1,\theta_2,\ldots,\theta_r$ are independent over $\Q$ if, considered as
vectors in the vector space $\R$ over the field $\Q$, they are linearly independent.
Equivalently, this is the case if the equality $m_1\theta_1+m_2\theta_2+\ldots+
m_r\theta_r=0$ for integer $m_1,m_2,\ldots,m_r$ implies $m_1=m_2=\ldots=m_r=0$.
Considering the actual physical system (a collection of metallic grains),
it is reasonable to assume that the capacitors $C_{i}$ are random, 
so that generically they are independent over $\Q$. Without loss of generality we 
may rearrange the $K$ capacitors such
that $C_{0}=\max_{0\le i\le K-1} C_{i}$. It is also useful to 
divide all the capacitors by the largest one, so that the 
scaled capacitors $c_{i} \equiv C_{i}/C_{0}$ with
$1=c_{0}>c_{1},c_{2}\ldots,c_{K-1}$ are dimensionless.
Finally, set $s=c_0+c_1+\ldots+c_{K-1}$.

Now we formulate our main results.

\proclaim Theorem 1. Suppose $C_0,C_1,\ldots,C_{K-1}$ are independent over $\Q$.
Then the sequence $(\chi_N)_{N=1}^\infty$ is asymptotically $F$-distributed, where
the distribution $F$ is given by either of the following two representations:
\begin{eqnarray}
F(x)&=&\cases{
               0,&$\ x<0,$\cr
               \displaystyle{1-\frac{1}{s}\sum_{i=0}^{K-1} c_i\prod_{j=0\atop j\ne i}^{K-1}
                   \left(1-\frac{x}{2w_j}\right),}&$\ 0\le x<2w_0,$\cr
               1,&$\ 2w_0\le x,$\cr
              }\label{eq_P_1(x)}\\\cr\cr\cr
    &=&\cases{
               0,&$\ x<0,$\cr
               \displaystyle{1-\frac{1}{s}\sum_{S\subseteq\{1,\ldots,K-1\}}}
                  \left(|S|+1\right)\prod_{i\in S} c_i
                  \prod_{i\notin S} (1-c_i)
                  \cdot\left(1-\frac{x}{2w_0}\right)^{|S|},&$\ 0\le x<2w_0,$\cr
               1,&$\ 2w_0\le x.$\cr
              }
\label{eq_P_2(x)}
\end{eqnarray}

It is not immediately obvious from the formulas, but $F$ has one discontinuity, namely
at the point $2w_0$. The reason is that, as the elements of $\Delta$ are all odd
multiples of the $w_i$'s, and as $w_0$ is the smallest of the $w_i$'s, it happens
occasionally that there is no odd multiple of $w_1,\ldots,w_{K-1}$ between two
consecutive multiples of $w_0$. The size of the atom at $2w_0$ is
$\frac{1}{s}\cdot\prod_{i=1}^{K-1} (1-c_i)$. This is easily explained intuitively.
In fact, the
``density" of odd multiples of $w_i$ is $c_i$ times the same density for multiples of
$w_0$. Hence the ``probability" that an interval of the form $[(2n-1)w_0,(2n+1)w_0)$
does not contain an odd multiple of $w_i$ is $1-c_i$. Assuming that the ``events"
of containing different $w_i$'s are independent, we conclude that the proportion
of multiples of $w_0$ in $\Delta$ whose successors are also such is
$\prod_{i=1}^{K-1} (1-c_i)$. Since the proportion of multiples of $w_0$ in $\Delta$
is $\frac{1}{s}$, we arrive at the required expression for the size of the atom.

Now we would like to study the asymptotic of the distances between consecutive
elements of $\Delta$ as the number of capacitors grows. Obviously, as this happens,
the distances become smaller. More precisely, on the average we have
$\frac{1}{2w_j}$ odd multiples of each $w_j$ in each unit interval, and hence we have
there $\sum_{j=0}^{K-1}\frac{1}{2w_j}=\frac{s}{2w_0}$ elements of $\Delta$ altogether.
Hence the average distance between consecutive elements is $\frac{2w_0}{s}$. To
understand the asymptotics of 
the gaps, it makes sense therefore to normalize them
so as to have mean 1. Thus, we multiply the distances by $\frac{s}{2w_0}$, and ask
about the asymptotic behavior

\proclaim Theorem 2. Suppose the capacitances $C_0,C_1,\ldots$ are chosen uniformly
and independently in $[0,1]$. For each $K$, let $F_K$ denote the distribution
corresponding to the normalized gaps when taking into account the first $K$
capacitors only. Then, with probability 1, the distributions $F_K$ converge
to an $\Exp(1)$ distribution function.

\demo Remark. As will be seen in the proof, we actually use much less to prove
Theorem 2 than is required by the conditions of the theorem. Namely, we need
the capacitances $C_i$ to be linearly independent over $\Q$, and that they do
not form a fast diminishing sequence.

It is worthwhile mentioning that this type of ``Poissonian" asymptotic behavior 
of consecutive gaps is typical. For example, this is the case for uniformly selected
numbers in $[0,1]$, and is conjectured to be the case in other interesting cases
as well. (See, for example, \cite{KR} and \cite{RZ} and the references there.)

In the course of the proof, we shall make use of the notion of uniform distribution
modulo~1 and a few basic results relating to it. (The reader is referred to
Kuipers and Niederreiter \cite{KN} for more information.) A sequence
$(x_n)_{n=1}^\infty$ of real numbers is {\it uniformly distributed modulo}~1 if
$$\frac{\left|\{1\le n\le M:a\le \{x_n\}<b\}\right|}{M}\longto_{M\to\infty}
        b-a,\qquad 0\le a<b\le 1\;,$$
where $\{t\}$ is the fractional part of a real number $t$.
In terms of Definition 1, $(x_n)$ is uniformly distributed modulo~1 if and
only if the sequence $(\{x_n\})$ of fractional parts is $F$-distributed, where
$F$ is the distribution function of the uniform distribution on $[0,1]$:
$$F(x)=\cases{
                        0,&$x<0,$\cr
                        x,&$0\le x\le 1,$\cr
                        1,&$x>1.$\cr
              } $$
The generalization of the notion of an asymptotically $F$-distributed sequence
to that of an asymptotically well $F$-distributed sequence clearly carries over
to our case. Instead of requiring only that the dispersion of large initial
pieces of the sequence becomes more and more even, we require this to
happen at arbitrary locations. This version is termed
{\it well-distribution}. Thus, $(x_n)_{n=1}^\infty$ is
{\it well-distributed modulo}~$1$ if
$$\frac{\left|\{L<n\le M:a\le \{x_n\}<b\}\right|}{M-L}\longto_{M-L\to\infty}
        b-a,\qquad 0\le a<b\le 1\;.$$
Both notions have multi-dimensional analogue. A sequence $(\x_n)_{n=1}^\infty$ in
$\R^s$ is {\it uniformly distributed modulo~1 in} $\R^s$ if
$$\frac{\left|\{1\le n\le N:\bfa\le \{\bfx_n\}<\bfb\}\right|}{N}\longto_{N\to\infty}
        \prod_{i=1}^s (b_i-a_i),\qquad \0\le \bfa<\bfb \le \1\;,$$
where inequalities between vectors in $\R^s$ are to be understood component-wise,
$\0=(0,0,\ldots,0)\in\R^s,\ \bfa=(a_1,a_2,\ldots,a_s)$, and so forth.

Perhaps the most basic example of a sequence which is uniformly distributed modulo~1
is $(n\alpha)_{n=1}^\infty$, where $\alpha$ is an arbitrary irrational. In the
multi-dimensional case, the sequence $(n\alpha_1,n\alpha_2,\ldots,n\alpha_s)$
is uniformly distributed modulo~1 in $\R^s$ if and only if the numbers
$1,\alpha_1,\alpha_2,\ldots,\alpha_s$ are linearly independent over $\Q$.
Moreover, in this case uniform distribution implies well-distribution
(cf. \cite[Example 1.6.1, Exercise 1.6.14]{KN}).

Given a partition $\N=\bigcup_{j=1}^l A_j$ and positive integers $r_j,\;j=1,\ldots,l$,
we define the $(r_j)_{j=1}^l$-{\it inflation} of the given partition as the partition
of $\N$ obtained by inflating each element of each of the sets $A_j$ into $r_j$
elements. More precisely, we construct sets $B_j,\;j=1,\ldots,l$, as follows.
For a positive integer $i$, let $f(i)=j$ if $i\in A_j$. Given
any positive integer $n$, let $m$ be defined by
$\sum_{i=1}^{m-1} f(i)< m\le \sum_{i=1}^{m} f(i)\;.$ Let $n\in B_j$ if $m\in A_j$.
The following lemma is routine.

\proclaim Lemma 2. In this setup:
\ben
\item
If $\D(A_j)=d_j,\ 1\le j\le l$, then $\D(B_j)=\frac{r_j d_j}{\sum_{i=1}^l r_i d_i}\;.$
\item
If $\BD(A_j)=d_j,\ 1\le j\le l$, then $\BD(B_j)=\frac{r_j d_j}{\sum_{i=1}^l r_i d_i}\;.$
\een

\bdemo Proof of Theorem 1. Between any two consecutive odd multiples of $w_0$, there
is at most one odd multiple of each $w_j,\ 1\le j\le K-1$. In fact, one easily verifies
that, given a positive integer $m$, there is an odd multiple of $w_j$ between
$(2m-1)w_0$ and $(2m+1)w_0$, namely there exists an integer $n$ with
\begin{eqnarray}
(2m-1)w_0\le (2n-1)w_j<(2m+1)w_0,\label{contain_w_j}
\end{eqnarray}
if and only if
\begin{eqnarray}
m c_j\in\left({1-c_j\over 2},{1+c_j\over 2}\right]\;(\mod 1)\;.\label{cond_contain_w_j}
\end{eqnarray}
Moreover, the relative position of $(2n-1)w_j$ within the interval
$[(2m-1)w_0,(2m+1)w_0)$ is the same, but in the opposite direction, as that of
$m c_j(\mod 1)$ within the interval $\big({1-c_j\over 2},{1+c_j\over 2}\big]$, that is
\begin{eqnarray}
(2n-1)w_j=\alpha\cdot(2m-1)w_0+(1-\alpha)\cdot(2m+1)w_0,\qquad (0<\alpha\le 1)\;,
\label{position_w_j}
\end{eqnarray}
if and only if
\begin{eqnarray}
m c_j\equiv (1-\alpha)\cdot{1-c_j\over 2}+\alpha\cdot{1+c_j\over 2}\;(\mod 1)\;.
\label{cond_position_w_j}
\end{eqnarray}

Next we define a partition of $\N$ as follows. Write the elements of $\Delta$
in ascending order:
$\Delta=\{\delta_1<\delta_2<\delta_3<\ldots\}$. Given $n\in\N$, let
$S\subseteq\{1,2,\ldots,K-1\}$ denote the set of all those $j$'s such that the unique
interval of the form $[(2m-1)w_0,(2m+1)w_0)$ containing $\delta_n$ contains an odd
multiple of $w_j$. The set of all integers $n$ giving rise in this way to any set $S$
is denoted by $B_S$. Consider the partition
$\N=\bigcup_{S\subseteq\{1,2,\ldots,K-1\}} B_S$.
To prove the theorem using Lemma~1, we have to find the Banach densities of the sets
$B_S$ and the asymptotic distribution of the corresponding subsequences
$(\chi_n)_{n\in B_S}$ of $\chi_n$.

The partition of $\N$ into sets of the form $B_S$ is obtained as an inflation of a
somewhat more straightforward partition. In fact, let $S$ be any subset of
$\{1,2,\ldots,K-1\}$. Denote by $A_S$ the set of those positive integers $n$ for which
the interval $[(2n-1)w_0,(2n+1)w_0)$ contains odd multiples of $w_j$ for $j\in S$ and
does not contain such multiples of the other $w_j$'s. Then
$\N=\bigcup_{S\subseteq\{1,\ldots,K-1\}} A_S$ is a partition, and its
$(|S|+1)_{S\subseteq\{1,2,\ldots,K-1\}}$-inflation yields the partition
$\N=\bigcup_{S\subseteq\{1,2,\ldots,K-1\}} B_S$.

In view of the equivalence of (\ref{contain_w_j}) and (\ref{cond_contain_w_j}), $A_S$
is the set of those $n$'s for which $n c_j\in\left({1-c_j\over 2},{1+c_j\over 2}\right]$
for $j\in S$ and $n c_j\notin\left({1-c_j\over 2},{1+c_j\over 2}\right]$ for $j\notin S$.
By the conditions of the theorem, the numbers $1,c_1,\ldots,c_{K-1}$ are linearly
independent over $\Q$, and hence the sequence
$\c=(nc_1,nc_2,\ldots,nc_{K-1})_{n=1}^\infty$
is well-distributed modulo~1 in $\R^{K-1}$. This means that
\begin{eqnarray}
\D(A_S)=\BD(A_S)=\prod_{i\in S} c_i \prod_{i\notin S} (1-c_i)\;.\label{density_A_S}
\end{eqnarray}
Denote the right hand side of (\ref{density_A_S}) by $p_S$.
In view of the above and Lemma 2, this implies
\begin{eqnarray}
\D(B_S)=\BD(B_S)=\frac{(|S|+1) p_S}{\sum_{T\subseteq\{1,2,\ldots,K-1\}} (|T|+1) p_T}\;.
\label{density_B_S}
\end{eqnarray}
The denominator on the right hand side can be given a simpler form. In fact, let
$X_i,\ i=1,2,\ldots,K-1,$ be independent random variables with $X_i\sim B(1,c_i)$,
and $X=\sum_{i=1}^{K-1} X_i$. Then:
\begin{eqnarray}
\sum_{T\subseteq\{1,2,\ldots,K-1\}} (|T|+1) p_T=E(X+1)=1+c_1+\ldots+c_{K-1}=s\;.
\end{eqnarray}
Hence:
\begin{eqnarray}
\BD(B_S)=\frac{(|S|+1) p_S}{s}\;.
\label{density_B_S_2nd_form}
\end{eqnarray}

Let $S$ be an arbitrary fixed subset of $\{1,2,\ldots,K-1\}$, say $S=\{1,2,\ldots,l\}$,
where $0\le l\le K-1$. If $n\in A_S$, then there exist odd integers
$a_{1n},a_{2n},\ldots,a_{ln}$ such that $a_{jn}w_j\in [(2n-1)w_0,(2n+1)w_0)$. Put:
$$v_n=(a_{1n}w_1,a_{2n}w_2,\ldots,a_{ln}w_l)-(2n-1)w_0\cdot(1,1,\ldots,1)
  \in [0,2w_0)^l,\qquad n\in A_S\;.$$
By the equivalence of (\ref{position_w_j}) and (\ref{cond_position_w_j}), the sequence
$(v_n)_{n\in A_S}$ is well-distributed modulo~$2w_0$ in~$\R^l$. Now each $v_n$ gives
rise to $l+1$ terms of $(\chi_n)_{n\in B_S}$, as follows. Let
$v_n^{(1)}\le v_n^{(2)}\le\ldots\le v_n^{(l)}$ be all coordinates of $v_n$ in
ascending order. Set:
$$u_n=(v_n^{(1)},v_n^{(2)}-v_n^{(1)},\ldots,v_n^{(l)}-v_n^{(l-1)},2w_0-v_n^{(l)}),
  \qquad n\in A_S\;.$$
The sequence $(\chi_n)_{n\in B_S}$ consists of all coordinates of all vectors
$u_n$. Now we use the fact that if $X_1,X_2,\ldots,X_r$ are independent random variables,
distributed ${\rm U}(0,h)$, and $X^{(1)},X^{(2)},\ldots,X^{(r)}$ are the corresponding
order statistics, then each of the random variables
$X^{(1)},X^{(2)}-X^{(1)},\ldots,X^{(r)}-X^{(r-1)},h-X^{(r)}$ has the distribution
function defined by
$G(x)=1-(x/h)^r$ for $0\le x\le h$ (which follows as a special case from
\cite[p.42, ex.23]{Fe}). Consequently, for each $1\le j\le l+1$, the sequence given
by the $j$th coordinate of all vectors $u_n,\ n\in A_S$, is asymptotically well
$G_1$-distributed, where $G_1(x)=1-(x/2w_0)^l$ for $0\le x\le 2w_0$. Hence the
sequence $(\chi_n)_{n\in B_S}$ is asymptotically well $G_1$-distributed. Combined
with (\ref{density_B_S_2nd_form}), it proves (\ref{eq_P_2(x)}).

We shall indicate only briefly the proof of (\ref{eq_P_1(x)}), which is quite
simpler. This time, we split $(\chi_n)$ into a union of subsequences
$(\chi_n^{(i)}),\ 0\le i\le K-1$, by putting $\chi_n$ in the sequence
$\chi_n^{(i)}$ if $\delta_n$ is a multiple of $w_i$. Clearly, the proportion
of terms of $(\chi_n)$ belonging to $(\chi_n^{(i)})$ is $c_i/s$. Next, consider
the minimal odd multiples of all $w_j$'s which are larger than $\delta_n$. The
minimum of these $K$ numbers is $\delta_{n+1}$. For each $j\ne i$, the distance from
$\delta_n$ to the minimal odd multiple of $w_j$ following $\delta_n$ is ``distributed"
${\rm U}(0,2w_j)$. (For $i=0$ it is also possible that the next term will be again a
multiple of $w_0$.) The linear independence of the $C_i$'s over $\Q$ implies that these
$K-1$ distances are (statistically) independent, so that their minimum is distributed
according to the function
$G_2(x)=1-\prod_{j=0\atop j\ne i}^{K-1}\left(1-\frac{x}{2w_j}\right)$
on the interval $[0,2w_0)$. These considerations can be formalized to prove
(\ref{eq_P_1(x)}). This completes the proof.
\edemo

\demo Remark. It is possible to shorten the proof by proving directly the equality
of the right hand sides of (\ref{eq_P_1(x)}) and (\ref{eq_P_2(x)}).
In fact, it is easy to integrate both forms with respect to $x$; the
equality of the resulting expressions follows easily from the binomial theorem.
We have chosen the long way, as it is more instructive.

\bdemo Proof of Theorem 2. The distribution $F_K$ is obtained from that in Theorem~1
by stretching by the constant factor $\frac{s}{2w_0}$. Hence:
\begin{eqnarray}
F_K(x)&=&\cases{
               0,&$\ x<0,$\cr
               \displaystyle{1-\frac{1}{s}\sum_{i=0}^{K-1} c_i\prod_{j=0\atop j\ne i}^{K-1}
                   \left(1-\frac{c_j x}{s}\right),}&$\ 0\le x<s,$\cr
               1,&$\ s\le x\;.$\cr
              }
\label{eq_F_K(x)}
\end{eqnarray}
Note that some of the values appearing on the right hand side depend on $K$ implicitly.
Namely, since $w_0$ is assumed in Theorem~1 to be the least $w_i$, each time a $C_i$
is selected which is larger than all the heretofore selected $C_j$'s, we have to
rearrange the $C_j$'s, thus changing $w_0$ and the $c_j$'s. We have to show that
\begin{eqnarray}
F_K(x)\longto_{K\to\infty} 1-e^{-x},\qquad x\ge 0\;.
\label{conv_F_K(x)}
\end{eqnarray}
Indeed, fix $x\ge 0$. Since
\begin{eqnarray}
s=c_0+c_1+\ldots+c_{K-1}=\frac{C_0+C_1+\ldots+C_{K-1}}{C_0}\ge C_0+C_1+\ldots+C_{K-1}
\end{eqnarray}
and the $C_i$'s are independent and uniformly distributed in $[0,1]$, we have
\begin{eqnarray}
s\longto_{K\to\infty}^{\rm a.s.}\infty\;.
\label{conv_s}
\end{eqnarray}
Hence, with probability 1, for sufficiently large $K$ we have
\begin{eqnarray}
F_K(x)=\displaystyle{1-\frac{1}{s}\sum_{i=0}^{K-1} c_i\prod_{j=0\atop j\ne i}^{K-1}
                   \left(1-\frac{x}{2w_j}\right)}\;.
\end{eqnarray}
Thus, to prove (\ref{conv_F_K(x)}) we need to show that
\begin{eqnarray}
\frac{1}{s}\sum_{i=0}^{K-1} c_i\prod_{j=0\atop j\ne i}^{K-1}
\left(1-\frac{c_j x}{s}\right) \longto_{K\to\infty}^{\rm a.s.} e^{-x},\qquad x\ge 0\;.
\end{eqnarray}
Now, on the one hand, using the inequality
$$1-t\le e^{-t},\qquad t\in\R\;,$$
we have
$$\prod_{j=0\atop j\ne i}^{K-1}\left(1-\frac{c_j x}{s}\right)\le
e^{-x\sum_{j=0\atop j\ne i}^{K-1}\frac{c_j}{s}}\le
e^{-x+x/s},\qquad i=0,1,\ldots,K-1\;,$$
and therefore
\begin{eqnarray}
\frac{1}{s}\sum_{i=0}^{K-1} c_i\prod_{j=0\atop j\ne i}^{K-1}
\left(1-\frac{c_j x}{s}\right) \le \frac{1}{s}\sum_{i=0}^{K-1} c_i e^{-x+x/s}
=e^{-x+x/s}\longto_{K\to\infty}^{\rm a.s.} e^{-x}\;.
\label{upper_bound}
\end{eqnarray}
On the other hand, as $t\to 0$ we have
$$e^{-(t+t^2)}=1-(t+t^2)+\frac{(t+t^2)^2}{2}+O(t^3)=1-t-\frac{t^2}{2}+O(t^3)\;,$$
so that for all $t$ in some sufficiently small neighborhood of $0$
$$e^{-(t+t^2)}\le 1-t\;.$$
Consequently:
\begin{eqnarray}
\prod_{j=0\atop j\ne i}^{K-1}\left(1-\frac{c_j x}{s}\right)\ge
e^{-x\sum_{j=0\atop j\ne i}^{K-1}\frac{c_j}{s}
   -x^2\sum_{j=0\atop j\ne i}^{K-1}\frac{c_j^2}{s^2}}\ge e^{-x-K x^2/s^2}\;.
\label{lower_bound_1}
\end{eqnarray}
Obviously, with probability 1, $s$ grows linearly with $K$, namely for
all sufficiently large $K$ we have $s\ge aK$ for a suitably chosen $a>0$.
(In fact, any $a<\frac{1}{2}$ will do.) By (\ref{lower_bound_1}):
\begin{eqnarray}
\prod_{j=0\atop j\ne i}^{K-1}\left(1-\frac{c_j x}{s}\right)\ge
e^{-x-K x^2/s^2}\longto_{K\to\infty}^{\rm a.s.} e^{-x}\;.
\label{lower_bound_2}
\end{eqnarray}
From (\ref{upper_bound}) and (\ref{lower_bound_2}) it follows that
\begin{eqnarray}
\frac{1}{s}\sum_{i=0}^{K-1} c_i\prod_{j=0\atop j\ne i}^{K-1}
\left(1-\frac{c_j x}{s}\right)
\longto_{K\to\infty}^{\rm a.s.} e^{-x}\;,
\end{eqnarray}
which completes the proof.
\edemo
A plot of $P(\chi)$ is given in figure ~~\ref{fig1}
 and 
shows indeed that it has a maximum at $\chi=0$ and a 
delta function component at the inverse of the largest 
capacitance. In fact, 
when $K \rightarrow \infty$, the weight of the 
delta function shrinks to zero and $P(\chi)$ approaches the 
Poisson distribution $e^{- \chi}$. \\ 
\noindent
\section{Fluctuation of Coulomb Blockade Peak Spacings
in Large Semiconductor Quantum Dots}
\subsection{Motivation}
Recently, it became apparent that 
the physics exposed in the addition spectra of quantum dots is 
rather rich, and hence its investigation is
at the focus of both experimental and theoretical studies. 
The present section concentrates on the distribution of spacings 
between Coulomb blockade peaks in large semiconductor quantum dots.
Coulomb blockade is evidently one of the hallmarks of 
mesoscopic physics. The experimental achievement of tracing an 
addition of a single electron to a quantum dot and the appearance 
of isolated conductance peaks led to the concept of 
single electron transistors. After 
the origin of Coulomb blockade peaks has been elucidated, investigation 
is directed toward more subtle questions like their heights, widths 
and spacings. The underlying physics is related 
to the ground state energy, chemical potential 
and inverse compressibility of a few electron island 
coupled capacitively to its environment, 
as well as fluctuations of these quantities
with the number $N$ of electrons on the dot. 

\noindent
As far as the distribution of spacings between adjacent
Coulomb blockade peaks is concerned, the 
question can be stated as follows: 
According to the simplest picture in which the 
quantum dot is regarded as a single electron island 
whose coupling with the leads is through 
its capacitance $C$, 
the total potential energy of a 
quantum dot is $Q^2/2C-V_g Q$ where $V_g$ is 
the corresponding gate voltage. The conductance peaks 
occur at those values of $V_g$ for which $C V_{g}=e (N+1/2)$ 
where $e$ is the electron charge (henceforth $e=-1$).
For this value of $V_g$ the
addition of an electron to the dot (which contains 
$N$ electrons) does not cost any 
charging energy.
The position of the $N^{th}$ Coulomb blockade peak is then a linear 
function of $N$ and therefore, 
the spacing should be a constant $1/C$, independent 
of $N$. Recent experiments \cite{Sivan,Kouenhoven} indicate 
however that spacing between Coulomb blockade peaks in small 
quantum dots is in general not constant but, rather, a fluctuating 
quantity close to Gaussian. The average of its distribution approximately 
coincides with the constant value mentioned above,
but the elucidation of its standard deviation is 
still under investigation \cite{Mirlin,Orgad}. 

\noindent
The situation is even less clear
if the quantum dot is very large. 
As indicated in a series of recent experiments, 
the spacing occasionally vanishes, 
namely, two peaks (and sometimes even three peaks) coincide. Moreover, 
the evolution of peak positions and spacings with an applied magnetic 
field indicates the existence of strong correlations between 
them \cite{Ashoori1,Ashoori2,Ashoori3}. 
These observation motivated numerous theoretical 
models based on the concept of pair tunneling \cite{Wan} 
or that of two-electron bound-states in depleted 
electron islands \cite{Raikh}.\\
\noindent
In the present section we examine the scenario 
according to which a large quantum dot 
like the one used in the last experiment \cite{Ashoori3}, is 
in fact, composed of {\em several} electron islands which are 
coupled capacitively among themselves as well as 
to the leads. (This is a natural extension of 
the scheme discussed in the first section 
in which no coupling is present). Electrons are added 
in such a way that the total potential 
energy of the dot is minimum. This simple generalization of 
the single island picture leads to a remarkable change in the 
spacing distribution from a Gaussian \cite{Sivan,Orgad} centered 
around a finite average to 
a one which is large at small spacings. When the coupling between 
islands is weak, the distribution has indeed a maximum at 
zero spacing. This result is short of explaining the perfect 
overlap of peaks, since it requires a delta function component
at zero spacing. Yet, it leads to the occurrence of couples and 
sometimes triples of closely spaced peaks, similar to the experimental
observation. Moreover, the evolution of the peak positions 
with the magnetic field is qualitatively similar to the 
experimental one. On the other hand, the present model does 
not predict a definite periodicity in the bunching of Coulomb 
blockade peaks with electron number $N$. 
In the next subsection the model is explained and 
the results of calculations are 
presented in the third subsection.
\section{Formalism}
Consider a large isolated two dimensional quantum dot 
in a perpendicular magnetic field $B$ subject to a gate 
voltage $V_{g}$. Unlike the traditional 
Coulomb blockade picture it might contain {\em several} 
electron islands which can be regarded as metallic 
objects with inductive couplings among themselves. 
These are determined by a positive 
definite symmetric matrix $C$ whose
diagonal elements $C_{ii} \equiv C_{i}>0$ are the 
corresponding capacities whereas the 
non-diagonal elements $C_{ij}=C_{ji}<0$ , ($i \ne j)$ are 
the corresponding coefficients of induction.
The electrostatic energy of such a system can be written as,
\begin{eqnarray}
E_{c}= \frac {1} {2} \sum_{i,j=1}^{K} p_{ij} N_{i} N_{j}- V_{g} N,
\label{eq_Ec}
\end{eqnarray}
where $N_{i}$ is the number 
of electrons on island $i$ 
(the number of islands $K>1$ might be around $10$), 
$N=\sum_{i=1}^{K} N_{i}$ is the 
total numbers of electrons, and the (symmetric positive-definite)
matrix $p \equiv C^{-1}$. 

\noindent 
Beside the electrostatic energy it is assumed that electrons in each island
occupy single particle quantum states (orbitals) 
whose energies $\epsilon_{i \alpha}$ ($i=1,2,..K; \alpha=1,2,..$) 
depend on the confining potential as well as on the
magnetic field. The latter is manifested
through its orbital effects as well 
as due to Zeeman splitting (in which case 
the quantum number $\alpha$ contains also a spin label). 
The corresponding occupation numbers 
$n_{i \alpha}$ can be either $0$ or $1$. 
The system described above might then be represented by
a classical Hamiltonian 
\begin{eqnarray}
H=H_{c} + H_{sp}, 
\label{eq_H}
\end{eqnarray}
where the charging Hamiltonian $H_{c}$ is just the electrostatic 
energy \ref{eq_Ec} written in terms of the orbital 
occupation numbers,
\begin{eqnarray}
H_{c}= \frac {1} {2} \sum_{i,j=1}^{K} p_{ij} 
[ \sum_{\alpha} n_{i \alpha} ]
[ \sum_{\alpha '} n_{j \alpha '} ]
-V_{g} \sum_{i=1}^{K} \sum_{\alpha} n_{i \alpha},
\label{eq_Hc}
\end{eqnarray}
and the single particle part of the Hamiltonian, $H_{sp}$  is,
\begin{eqnarray}
H_{sp}=\sum_{i=1}^{K} \sum_{\alpha} \epsilon_{i \alpha} n_{i \alpha} .
\label{eq_Hsp}
\end{eqnarray}
The precise form of the matrix elements $p_{ij}=[C^{-1}]_{ij}$
as well as the single particle energies $\epsilon_{i \alpha}$ 
are specified in the next section when we present our results.
Despite the fact that the Hamiltonian $H=H_{c} + H_{sp}$ is classical 
(and relatively simple) the elucidation of its spectrum
 for large $N$ and $K$ is virtually hopeless. In order 
to compute the ground state energy $E(N)$ 
one has to find the minimum of $H$ 
on all the possible sets ($N_{1},N_{2}...N_{K}$) with the 
constraint $\sum_{i=1}^{K} N_{i}=N$.  Note that 
the so called ``Coulomb Glass'' model obtains 
as a special case when $i$ refers 
to a lattice site with random energy 
$\epsilon_{i \alpha}=\epsilon_{i}$,
and a single orbital $N_{i}=n_{i}=0,1$. The interaction matrix is
then given by $p_{ij}=1/r_{ij}$ for $i \ne j$ and $p_{ii}=0$ where 
$r_{ij}$ is the distance between sites $i$ and $j$. 

\noindent
The position of the $N^{th}$ conductance peak is given by the 
first difference of the ground state energy, namely, the 
chemical potential of the (isolated) dot, 
$\mu \equiv E(N+1)-E(N)$. The spacing between peaks is 
determined by the second difference (the inverse compressibility),   
defined already in equation \ref{eq_compress}

The occurrence of close peaks 
for certain values of electron number $N$ correspond 
to small values of $\chi_{N}$ (recall that
for a single island quantum dot in which the single 
particle energies are neglected,
the inverse compressibility is a constant 
$1/C_{11}$). For the more general model described above 
the spacing distribution will of course fluctuate. 
In general, some constants appearing in the 
Hamiltonian $H=H_{c}+H_{sp}$ are 
random, (e.g. the elements of the matrix $C$ and  the 
single particle energies $\epsilon_{i \alpha}$), 
but most experiments are performed on a single
quantum dot, so that fluctuations are meant with 
respect to the electron number $N$.
The numbers  $\chi (N)$ might then be
considered as values assumed by a random variable $\chi$
which has a certain distribution 
function $P(\chi)$. \\ 
\noindent
\section{Results}
We now return to the full Hamiltonian 
of Eq. \ref{eq_H}. It contains 
the coefficients of capacitance and induction
matrix $C_{ij}$ and the single particle energies 
$\epsilon_{i \alpha}$ as input. In choosing 
the actual numerical values we use a few 
guidelines, one of them is to avoid too many 
independent input data. As will be clear below, these are 
{\em not} fitting parameters but rather, a set of constants 
which are chosen once for all on general physical grounds.  
First, for the electrostatic part, recall that
the matrix $C$ should be a symmetric positive definite matrix 
with $C_{ii}>0$ and $C_{ij}<0$ for $i \ne j$. Besides, we 
expect it to be random. We then assume that $C_{ii}$ are 
random numbers uniformly distributed between $0$ and $W$ 
whereas the non-diagonal elements are uniformly distributed 
between $-w$ and $0$. Evidently, in choosing 
these constants, one has to 
keep $w << W$ in order to maintain positive 
definiteness. Actually, it is only the ratio $w/W$ which 
matters, so one can assume $W=1$ and use $e^{2}/W=1$ as 
an energy unit, leaving $w$ as a constant reflecting the 
strength of coupling between islands.
Second, for the single particle energies,
we consider each electron 
island $i$ as a two-dimensional potential well 
$V_{i}(r)=\frac {1} {2} M \omega_{i}^{2} r^{2}$, where 
$M$ is an effective mass. 
One may then regard $\omega_{i}^{-1/2}$ 
as a measure of the radius of the corresponding 
electron island. Since the capacitance $C_{ii}$ 
is also proportional to this radius we assume 
$\omega_{i}= \gamma  C_{ii}^{-2}$, where $\gamma$ is 
a constant reflecting the relation between 
charging energies and single particle energies. The single particle 
energies in each electron island (subject to a perpendicular 
magnetic field $B$)
are then known analytically. To be more specific recall that 
in two dimensions there are two quantum numbers for 
the orbital motion (denoted hereafter as $n,m$) to which 
we add a spin index $\sigma = \pm 1$. Then, 
with $\alpha =(n,m,\sigma)$, we have
\begin{eqnarray}
\epsilon_{i \alpha}= \frac {\hbar} {2} \large [ n \omega_{c} 
+ m \sqrt {\omega_{i}^{2}+\frac {1} {4} \omega_{c}^{2}} \large ] 
+ g \mu_{B} \sigma B,
\label{eq_eps} 
\end{eqnarray}
where $\omega_c=eB/Mc$ is the cyclotron frequency, $g$ is the 
g-factor and $\mu_B$ is the Bohr magneton. Since $g$ contains 
the effective mass it is not known accurately. 
Its value is constrained on physical grounds (see below). 
Note that in this scheme, the spacings between single particle 
energies are deterministic, and do not 
follow the Wigner surmise. The main cause of 
fluctuation is then due to the combination of non-random 
single particle energies and the occurrence of
numerous charging energies. Finally, the 
number of electron islands $K$ 
is determined by the size of the quantum dot.
The four input data of the model are then $K,w,\gamma$ 
and $g$.
Note that the gate voltage $V_g$ does not have an important role 
here. Indeed,
in actual experiments the variation of gate voltage serves to 
adjust the energies $E(N)$ with the chemical potential of 
the leads but here the ground state energies are calculated directly. \\
\noindent
In order to avoid redundancy we are content with 
having a single set of these constants which is physically 
reasonable. In particular, it assures that the charging 
energy is much larger than single particle level spacings 
and that the Zeeman splitting is small at moderate magnetic fields.
Specifically, we take $w=0.03$, $\gamma=0.1$ and $g$
 is chosen such that for moderate fields the Zeeman 
splitting is of the order of the mean level spacing.\\
\noindent
With these prescriptions, we can
 find the 
ground state energy as a function of the 
magnetic field $B$ for each electron number $N$. 
We use just a brute force trial algorithm, 
and hence cannot treat systems with large number of electron 
islands. An intriguing question is whether the addition of 
an electron leads to a redistribution of all electrons among 
the islands or else, the newly added electron will chose a place 
such that the energy cost is minimal while all others remain 
intact. Our results indicate that in most cases there is no 
redistribution, but in some cases redistribution does occur 
although it does not involve an overall re-shuffling. \\ 
\noindent
Consider as a representative example 
a quantum dot with $K=5$ electron islands.  
The ground state energies $E(N)$ 
and the inverse compressibilities $\chi(N)$ 
(in the absence of a magnetic field) 
are calculated for electron 
number $N$ up to $200$. The first question we addressed is 
how the electrons are added among the islands. 
Figure ~~\ref{fig2} shows electron numbers $N_{i}$ in each one 
of the five islands, {\em v.s} the total electron number $N$, in 
the range $60<N<80$. Evidently, the order of curves is according 
to the value of the capacitance $C_{ii}$. On a larger scale, the 
numbers $N_{i}$ grow linearly with $N$ as it should be. Let us
then consider the question of redistribution. From figure ~~\ref{fig2} 
we see that redistribution occurs only once, as $N$ grows between 
$67$ and $68$ (see the vertical line). The addition of electron 
to the second (or third) island involves also a transfer of an 
electron from the fourth island to the second (or the third) one. 
This scenario occurs also in other ranges of $N$ with the same 
proportion (namely about four percents). Thus, within the present model, 
redistribution is present although it is rare and minimal.\\
\noindent
The next question is related to the fluctuation of the 
inverse compressibility $\chi (N)$. Figure ~~\ref{fig3} displays 
$\chi (N)$ {\em v.s} $N$ for $40<N<80$. From this figure 
one notices that the inverse compressibility fluctuates 
quite strongly. In particular, it becomes small, 
(compared with its average) although
it does not vanish. Note also that there is, at first 
glance, no traces of any periodic structure. This is also 
verified on applying its Fourier transform. The distribution 
$P(\chi)$ is drawn in figure ~~\ref{fig4}. It is indeed 
very different from the constant $P(\chi)=\delta(\chi-1/C)$ 
appropriate for a single island quantum dot with capacitive 
coupling $C$. A trace of this latter characteristic is 
the peak near $\chi=1$. The eventual decrease of the distribution 
near $\chi=0$ seems to be related to the fact that, unlike 
the case of independent systems, the capacitance matrix 
is non-diagonal. Switching on the coupling here then  
has an effect similar to a weak ``spacing repulsion'', 
similar to the  familiar effect of perturbation in
 two level systems. Recall,
however, that beside the capacitance induction matrix $C$ the 
total energy is determined also by the 
single particle energies in each island. The remarkable 
point is that the distribution is not Gaussian (which is 
the hallmark of spacing distributions in small quantum dots).\\
\noindent
Finally, we check the behavior of the first difference 
$E(N+1)-E(N)$ as function of the magnetic field. Recall that 
this quantity is proportional to 
the position of the $N^{th}$ Coulomb blockade 
peak. As a measure of the strength of the magnetic field we 
use the parameter $\omega_{c} / \omega_{0}$ where $\omega_{0}$ 
is the harmonic oscillator frequency of the largest island. 
The positions of the peaks for $39<N<48$ are displayed in 
figure ~~\ref{fig5}. Comparison with the results 
displayed in figure 2 of Ref. \cite{Ashoori3} reveals 
a large degree of qualitative agreement. In particular, 
it shows that groups of two (and sometimes even three) 
electrons can tunnel through the quantum dot at almost 
the same gate voltage. The oscillations at small 
magnetic field just mark transitions to lower Landau levels 
as the magnetic field increases. The phenomena of alternate 
bunching $(N,N+1) \rightarrow (N-1,N)$ is also reproduced 
in the present picture. \\
\noindent
In conclusion, we suggest a classical 
model in which a large  semiconductor quantum 
dot is viewed as a collection of metallic 
electron islands with capacitive and inductive 
coupling among them. The effect of magnetic field 
is manifested through its orbital as well as its 
spin effects. The model 
can explain the occasional occurrence of 
couples or even triples of closely spaced Coulomb blockade peaks, 
as well as the qualitative behavior of peak positions 
with the magnetic field. 
\begin{acknowledgments}
his work was supported in part by grants from the Israel Academy of
Science and Humanities under the program
{\em Centers for Excellence}, by the Basic
Research Foundation and the BSF -- Binational Israel-US Foundation.
One of us (Y. A) is grateful to 
R. Ashoori and D. V. Averin for discussion and suggestions.
\end{acknowledgments}

\newpage
\begin{figure}
\caption{Distribution of inverese compressibility $P(\chi)$ 
for a system containing independent subsystems each of 
which has a ground state energy proportional to the 
square of the number of particles it contains. The 
graph corresponds to a special case of equation 
\ref{eq_Pchi} with $K=5$.
\label{fig1}}
\end{figure}

\begin{figure}
\caption{
Distribution of electrons among the five islands 
as function of the total electron number $N$. 
The constants $K$, $w$, $\gamma$ and $g$ are as 
explained in the text.
The number $N_i$ of electrons on island $i$ is commensurate 
with its capacitance $C_{ii}$. In most cases the 
addition of an electron does not perturb the occupation 
of other islands. An example of redistribution is marked 
with a vertical line. An addition of an electron causes 
a minimal redistribution.
\label{fig2}
}
\end{figure}

\begin{figure}
\caption{
Inverse compressibility $\chi_N$ as function of 
electron number $N$ in the dot. The units on the ordinate 
are the charging energies of the latgest island, 
namely, $e^{2}/W=1$ (see text).
\label{fig3}
}
\end{figure}

\begin{figure}
\caption{Distribution $P(\chi)$ of level spacings in the dot 
at zero magnetic field (not normalized). Note the peak at 
small values of $\chi$.
\label{fig4}} 
\end{figure}

\begin{figure}
\caption{Coulomb blockade peak position for electron number 
$N$ between $39$ and $48$ as function of the magnetic field. 
Here $\omega_{o}$ is the oscillator frequency of the largest 
island and $\omega_c$ is the cyclotron frequency. The units on 
the ordinate are energy units as explained previously. They 
are proportional to the gate voltage appropriate 
for the corresponding peak.
\label{fig5}}
\end{figure}


\newpage
\begin{references}
\bibitem{Wigner} E. P. Wigner, Ann. Math. {\bf 53}, 36, (1951); {\bf 62},
548 (1955); {\bf 65}, 203 (1957); {\bf 67}, 325 (1958).
\bibitem{Dyson} F. J. Dyson, Jour. Math. Phys. {\bf 3}, 140 (1962);
{\bf 3}, 157 (1962); {\bf 3}, 166 (1962).
\bibitem{Bohigas}
 O. Bohigas, M.J. Giannoni and C. Schmit, Phys. Rev. Lett. 52, 1 (1984).
\bibitem{Altshuler} B. L. Altshuler and B. I. Shklovskii, Zh. Eksp.
Teor. Fiz. {\bf 91}, 220 (1986). [Sov. Phys. JETP {\bf 64}, 127 (1986)].
\bibitem{Sivan} U. Sivan, R. Berkovits, Y. Aloni, O. Prus, A. Auerbach 
and G. Ben Yoseph, Phys. Rev. Lett. {\bf 77}, 1123 (1996); 
F. Simmel, T. Heinzel, and D. A. Wharam, Europhys. Lett. {\bf 38},
123 (1997); S. R. Patel, 
S. M. Cronenwel, P. R. Stewart, A. G. Huiberg, C. M. Marcus, C. I. 
Durooz, J. S. Harris, K. C. Kampman and A. C. Gossard, Phys. Rev. 
Lett.m {\bf 80}, 4522 (1998).
\bibitem{ZAPW} N.B. Zhitenev, R.C. Ashoori, L.N. Pfeiffer and K.W. West,
{\it Phys. Rev. Lett.} {\bf 79}(1997), 2308.
\bibitem{Berry} M.V. Berry and M. Tabor, Proc. Roy. Soc. London 356, 375 (1977)\
.
\bibitem{Fe} W. Feller,
{\it An Introduction to Probability Theory and Its Applications}, Vol. II, 2nd \
ed.,
Wiley, New York, 1971.

\bibitem{Fu} H. Furstenberg,
{\it Recurrence in Ergodic Theory and Combinatorial Number Theory},
Princeton University Press, Princeton, New Jersey, 1981.
\bibitem{KN} L. Kuipers and H. Niederreiter,
{\it Uniform Distribution of Sequences},
Wiley, New York, 1974.
\bibitem{KR} P. Kurlberg and Z. Rudnick,
The distribution of spacings between quadratic residues, preprint.

\bibitem{RZ} Z. Rudnick and A. Zaharescu,
The distribution of spacings between 
small powers of a primitive root, preprint\
.
\bibitem{Kouenhoven} L. P. Kouenhoven, T. H. Osterkamp, M. W. S. 
Danoesastro, M. Eto, D. G. Austing, T. Honda and S. Tarucha, 
Science, {\bf 278}, 1788 (1997).
\bibitem{Mirlin} R. Berkovits and B. L. Altshuler, Phys. Rev. {\bf B55}, 
5297 (1997); Ya. M. Blanter, A. D. Mirlin and B. A. Muzikantskii, 
Phys. Rev. Lett. {\bf 78}, 2449 (1997); A. A. Koulakov, F. G. Pikus 
and B. I. Shklovskii, Phys. Rev. {\bf 54}, 9223 (1997).
\bibitem{Orgad} D. Orgad and S. Levit, unpublished.
\bibitem{Ashoori1} R. C. Ashoori {\em et al.}, Phys. Rev. Lett. {\bf 68}, 
3088 (1992).
\bibitem{Ashoori2} R. C. Ashoori {\em et al.}, Physica (Amsterdam) 
{\bf 189B}, 117 (1993).
\bibitem{Ashoori3} N.B. Zhitenev, R.C. Ashoori, L.N. Pfeiffer and K.W. West,
{\it Phys. Rev. Lett.} {\bf 79}(1997), 2308.
\bibitem{Wan} Y. Wan, G. Ortiz and P. Phillips, Phys. Rev. Lett. 
{\bf 75},2879 (1995); Phys. Rev. {\bf B 55}, 5313 (1997); Phys. 
Rev. Lett. {\bf 78}(C), 3979 (1998).
\bibitem{Raikh} M. E. Raikh, L. I. Glazman and L. E. Zhukov, 
Phys. Rev. Lett. {\bf 77}, 1354 (1996); Phys. Rev. Lett. 
{\bf 78}(C) 3980 (1998).
\bibitem{Berry} M. V. Berry and M. Tabor, Proc. Roy. Soc. London {\bf 356}
375 (1977)
\bibitem{JphysA} Y. Avishai, D. Berend and R. Berkovits,  
Journal of
Physics {\bf A} (Math. Gen.), in press (1998).
\end{references}
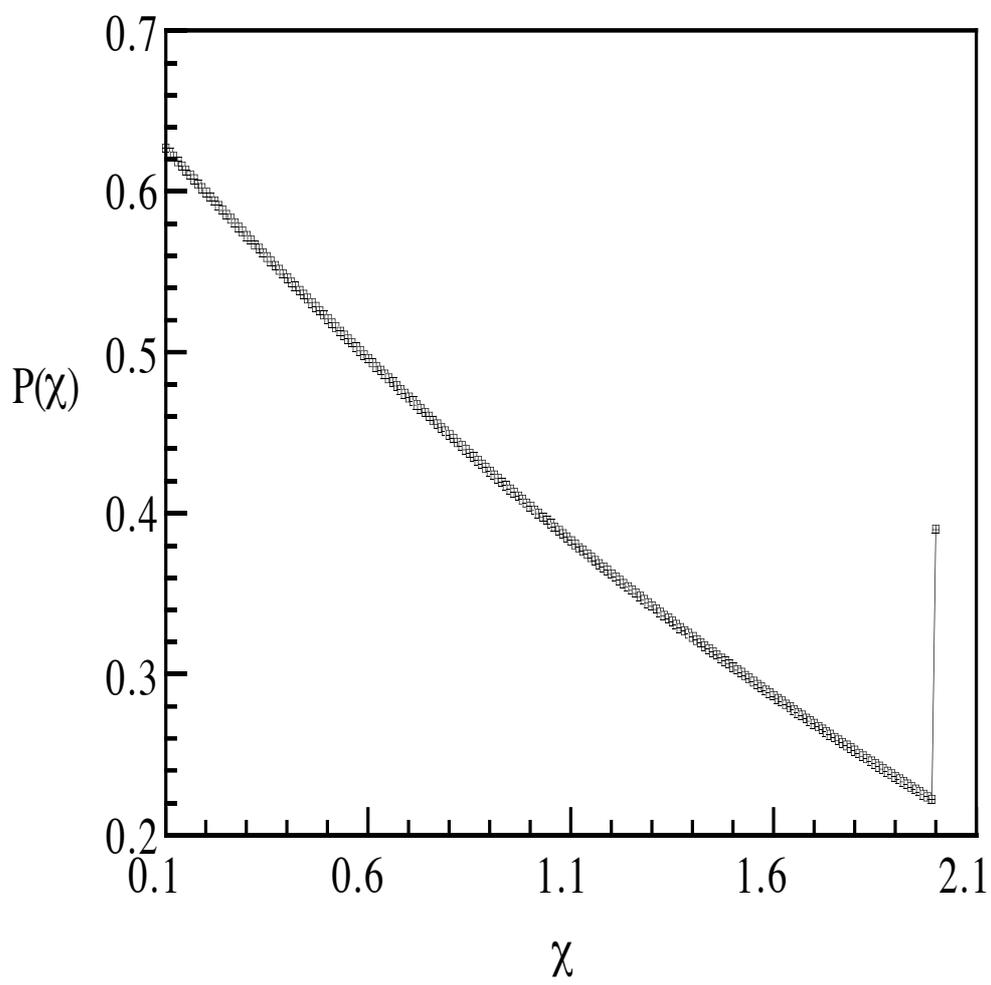
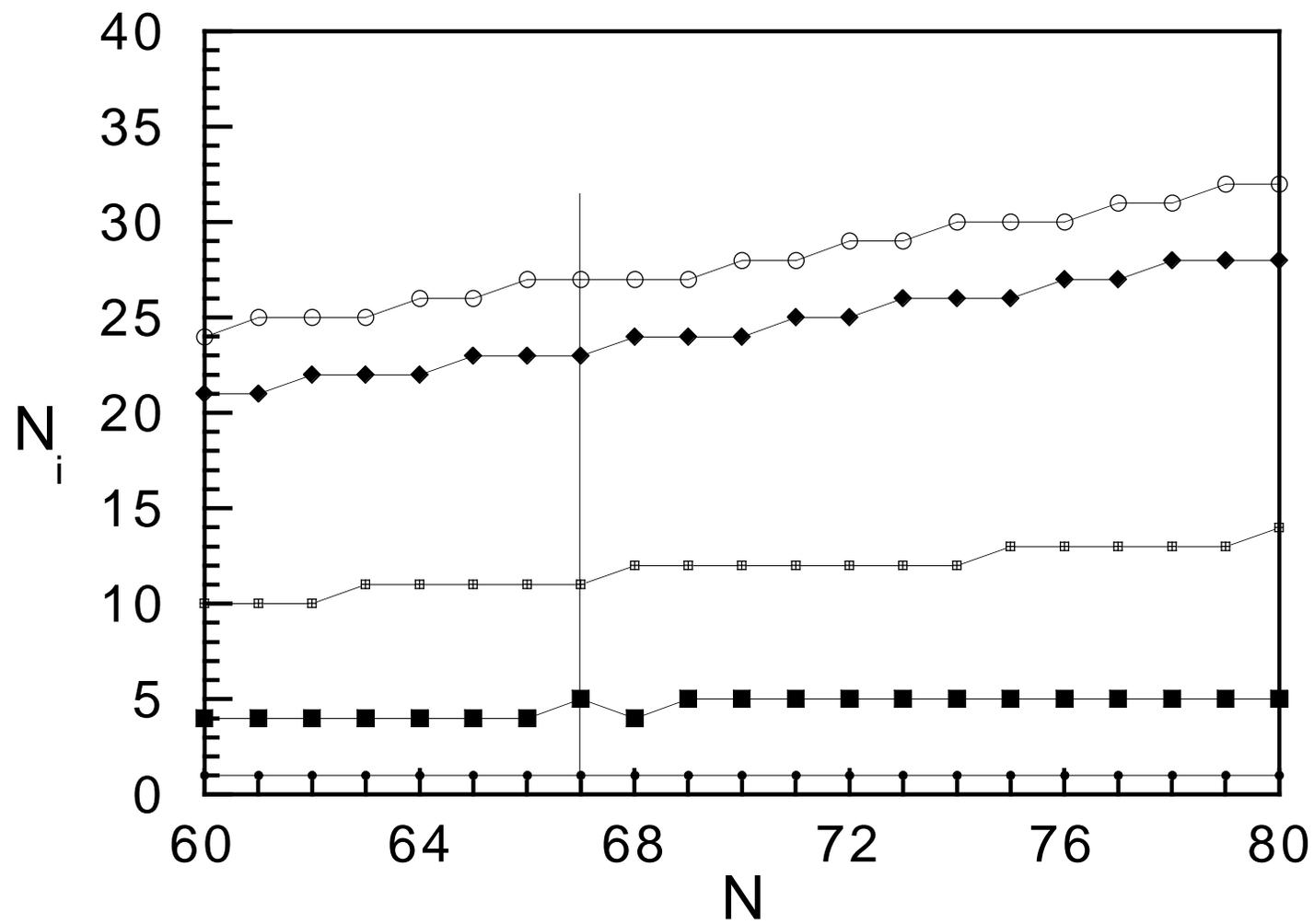
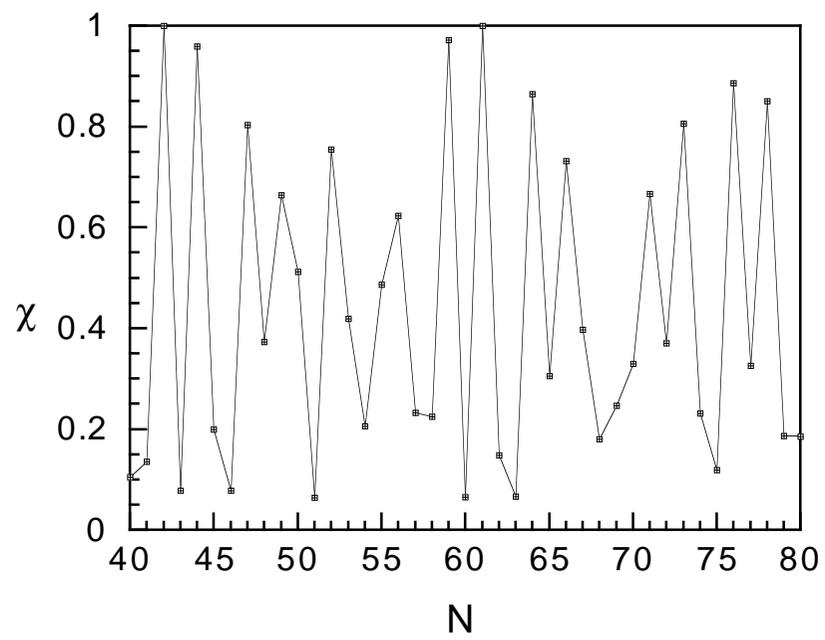
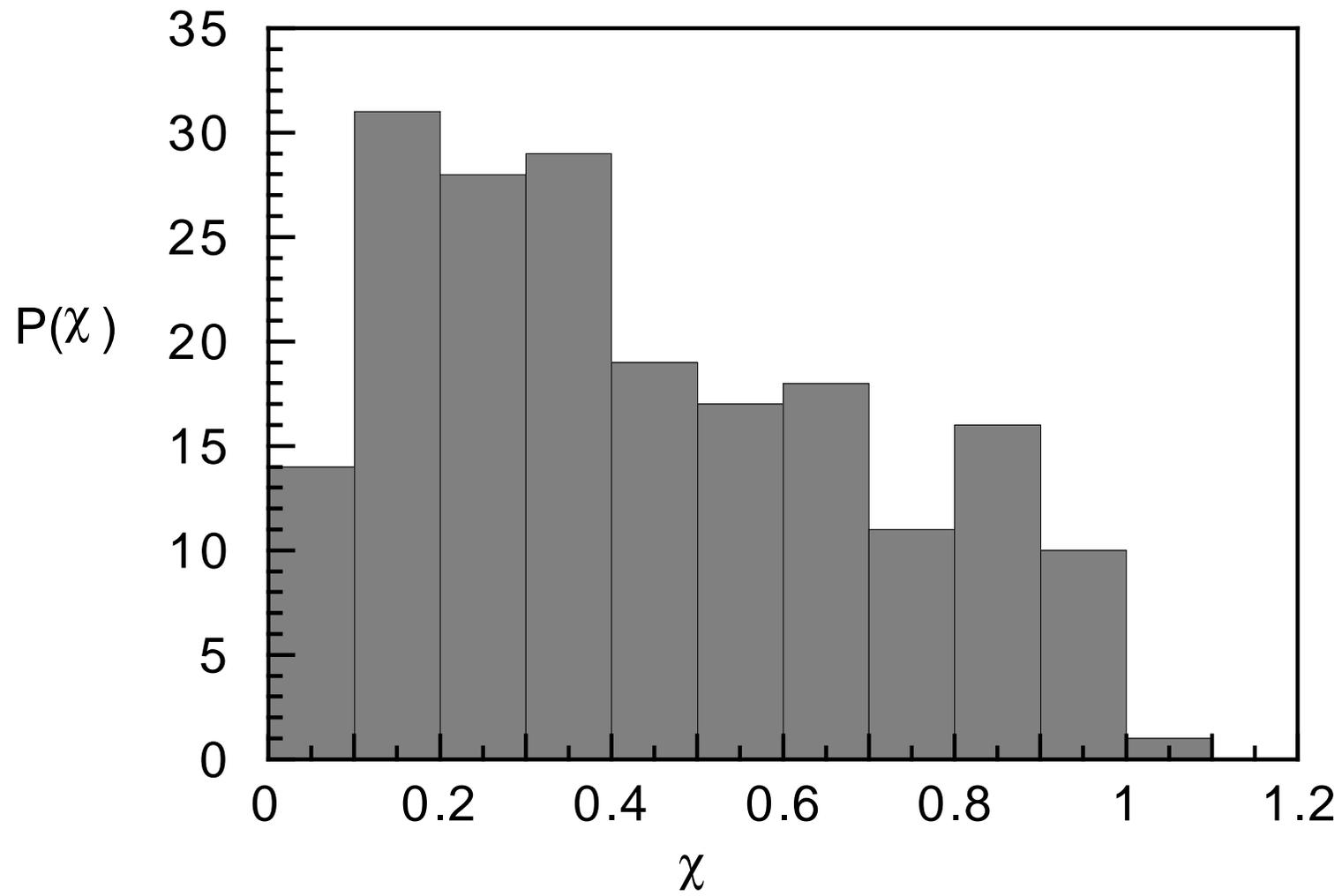
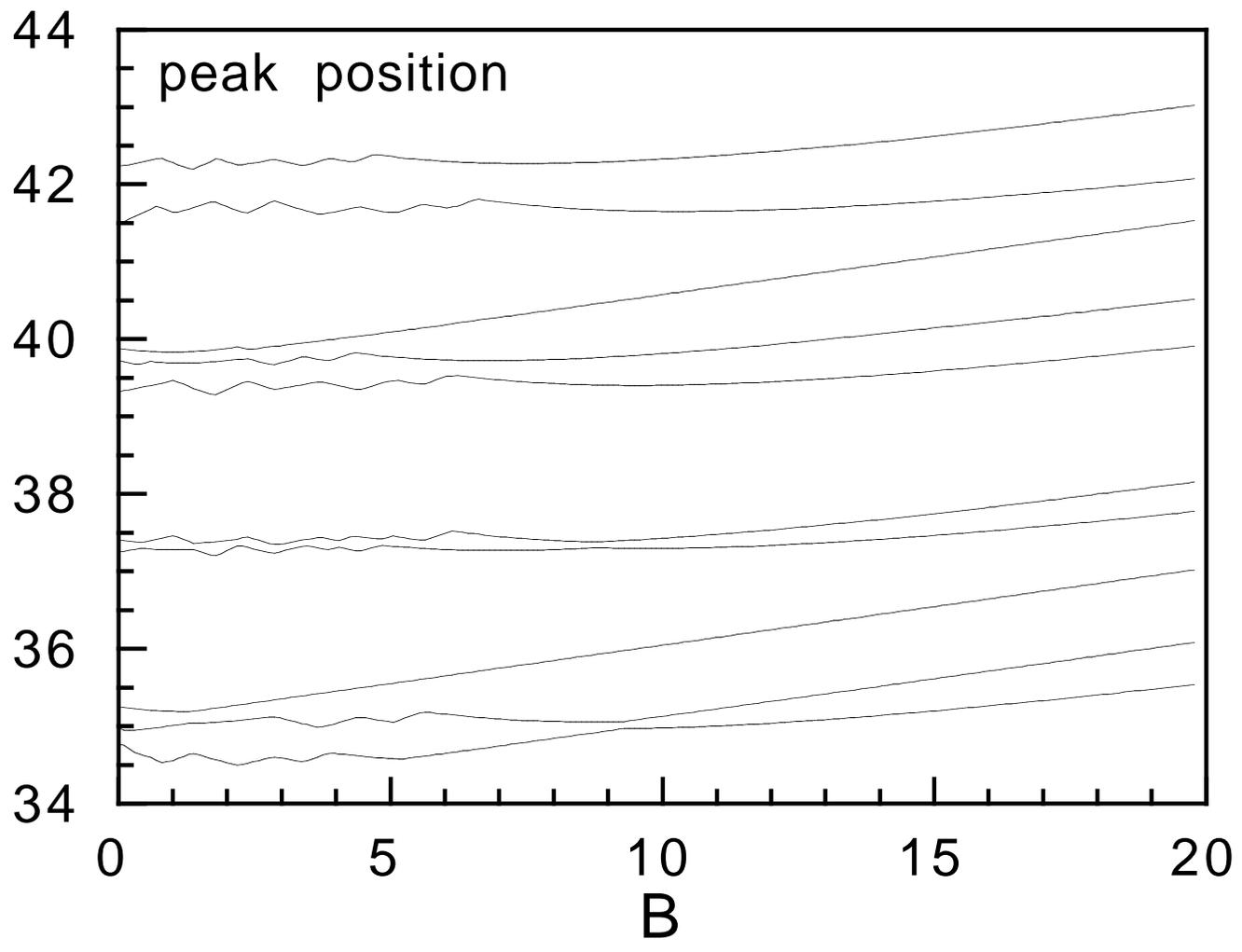
\end{document}